\documentclass{article}

\usepackage[english]{babel}
\usepackage[pdftex]{graphicx}
\usepackage{float}
\usepackage{subcaption}
\usepackage{amsmath}
\usepackage{amsfonts}
\usepackage{amsthm}
\usepackage{mathtools}
\usepackage{mathrsfs}
\usepackage[utf8]{inputenc}
\usepackage[top=1 in, bottom=1 in, left=1 in, right=1 in]{geometry}

\usepackage{color}
\usepackage{siunitx}
\usepackage{pgf}
\usepackage{svg}

\usepackage[
    natbib=true,
    backend=bibtex,
    style=nature,
    sorting=none,
    giveninits=true,
    uniquename=init
    ]{biblatex}

\usepackage{changes}
\usepackage[capitalise]{cleveref}

\newcommand{\rout}{r_\mathrm{out}}
\newcommand{\Iin}{I_\mathrm{in}}
\newcommand{\Vm}{V_\mathrm{m}}

\newcommand{\gNa}{g_\mathrm{Na}}

\newcommand{\gNabar}{\bar{g}_\mathrm{Na}}
\newcommand{\gKbar}{\bar{g}_\mathrm{K}}
\newcommand{\am}{\alpha_m}

\newcommand{\bm}{\beta_m}
\newcommand{\minf}{m_\infty}
\newcommand{\taum}{\tau_m}
\newcommand{\ah}{\alpha_h}
\newcommand{\bh}{\beta_h}

\newcommand{\hinf}{h_\infty}
\newcommand{\tauh}{\tau_h}
\newcommand{\an}{\alpha_n}
\newcommand{\bn}{\beta_n}
\newcommand{\Cm}{C_\mathrm{m}}
\newcommand{\gl}{g_\ell}

\newcommand{\El}{E_\ell}

\newcommand{\rhh}{r_\mathrm{HH}}
\newcommand{\rhhinf}{\rhh^\infty}
\newcommand{\rhhpros}{\rhh^\mathrm{pros}}
\newcommand{\trhh}{\tau_r^\mathrm{HH}}
\newcommand{\rthr}{r_\vartheta}
\newcommand{\rthrinf}{\rthr^\infty}
\newcommand{\rthrpros}{\rthr^\mathrm{pros}}
\newcommand{\trthr}{\tau_r^\vartheta}

\newcommand{\dt}{{\rm d}t}

\title{Prospective and retrospective coding in cortical neurons}

\author{Simon Brandt, \\ 
Mihai Alexandru Petrovici, Walter Senn, \\
Katharina Anna Wilmes*, Federico Benitez*\\
    \small{
       \textit{Department of Physiology, University of Bern, Switzerland}
        }\\\footnotesize{
        $^*$ Joint senior authorship
        }
}

\begin{document}

\maketitle

\begin{abstract}
Brains can process sensory information from different modalities at astonishing speed; this is surprising as the integration of inputs through the membrane of each individual neuron already causes a delayed response.
Neuronal recordings {\em in vitro} reveal a possible explanation for this fast processing, in terms of individual neurons advancing their output firing rates with respect to the input, a concept which we refer to as prospective coding.
The underlying mechanisms of prospective coding, however, are not completely understood.
We propose a mechanistic explanation for individual neurons advancing their output on the level of single action potentials and instantaneous firing rates.
We show that the spike generation mechanism can be the source for prospective (advanced) or retrospective (delayed) responses.
A simplified Hodgkin-Huxley model identifies sodium inactivation as a source for prospective firing, controlling the timing of the neuron's output as a function of the voltage and its temporal derivative.
We further show that slow adaptation processes, such as spike-frequency adaptation or deactivating dendritic currents, represent mechanisms generating prospective firing for inputs that undergo slow temporal modulations. In general, we show that adaptation processes at different time scales can cause advanced neuronal responses to time-varying inputs that are modulated on the corresponding time scales. 
\end{abstract}

\section*{Introduction}

Humans and other animals must adapt their behavior to a complex and dynamic world of physical and social interactions.
Adequately navigating our environments requires both fast processing of external information, e.g., in the form of accurate real-time sensory perception, and the anticipation of future states. 
Therefore, our brains evolved to process sensory information from different modalities at astonishing speed \citep{thorpe_speed_1996}. The brain must face the challenge of achieving this speed despite the fact that the integration of inputs through layers of neurons should cause a delayed response and hence slow signal processing. 

Prospective coding provides an elegant solution to this challenge.
We define prospective coding, or prospectivity, as neuronal firing being advanced with respect to the input. 
There is ample experimental evidence that the brain can process information prospectively in spite of the many forms of delays incurred by the underlying neuronal networks \citep{rainer_prospective_1999,ferbinteanu_prospective_2003}. However, a mechanistic explanation for how the brain achieves prospective coding is still missing. 

At the level of populations of neurons, it is a well-established fact that adaptation processes such as spike frequency adaptation can lead to prospective behavior \citep{puccini_integrated_2007}. Temporal processing in the form of advanced and delayed responses have also been observed to emerge from recurrent connectivity at the neuronal and population level \citep{dalla_porta_exploring_2019}.

Here, we consider prospective coding mechanisms taking place at the level of individual neurons, as opposed to being a collective phenomenon of populations of neurons \citep{abbott_building_2016}. In our approach, unlike \citep{brea_prospective_2016}, this individual prospectivity would be an intrinsic property of neurons, and not dependent on specific learning procedures.
Promisingly, prospective coding at the level of individual neurons has been observed experimentally, at least at the level of statistical ensembles.
In the study by Köndgen et al. \cite{kondgen_dynamical_2008} (see also \cite{linaro_dynamical_2018}), \emph{in vitro} pyramidal neurons were shown to be on average phase-advanced with respect to their noisy input.
This average advance occurs on short timescales, i.e.\ on the order of milliseconds.
Additionally, experiments have measured advanced responses on longer timescales, specifically in the case of neurons undergoing threshold adaptation \citep{fuhrmann_spike_2002,pozzorini_temporal_2013}. More generally, adaptation processes can occur simultaneously on different timescales, like fast sodium-channel inactivation coupled to slow after-hyperpolarization \citep{lundstrom_fractional_2008}.

On the computational side, cortical computation benefits from looking ahead in time to compensate neuronal delays, as shown by the recent development of prospective coding frameworks such as neuronal least action \cite{senn_neuronal_2024}, latent equilibrium \cite{haider_latent_2021}, and generalized latent equilibrium \cite{ellenberger_backpropagation_2024}.
All these methods consider prospective coding at the level of individual neurons, for instance with a rate that is not just a function of the voltage, but of the voltage plus its temporal derivative.
This linear look-ahead with the voltage-derivative allows individual neurons to advance their output firing rate with respect to their input.

In this work, we elucidate the cellular mechanisms underlying prospective coding on the individual neuron level. To this end, we use different biophysical models of cortical neurons to explore the circumstances under which neurons can advance their output firing rate relative to their input.
Our abstract model provides an overarching description of the biological phenomenon.
By simulating Hodgkin-Huxley-type neurons, we show how the action potential generation mechanism enables both advanced and delayed signals, reproducing experimental findings.
We show how these advances and delays can be observed for single neurons in single trials, excluding the possibility that the prospective and retrospective responses are a mere statistical effect. 
We reduce the Hodgkin-Huxley equations to a simple one-variable model that faithfully captures the prospective or retrospective firing. 
We further extend this approach to more general rate models, where adaptation processes can lead to prospective coding on the order of hundreds of milliseconds.

\section*{Results}
\subsection*{Hodgkin-Huxley neurons can fire prospectively and retrospectively}
\begin{figure}[H]
    \centering
    \includegraphics[width=1\linewidth]{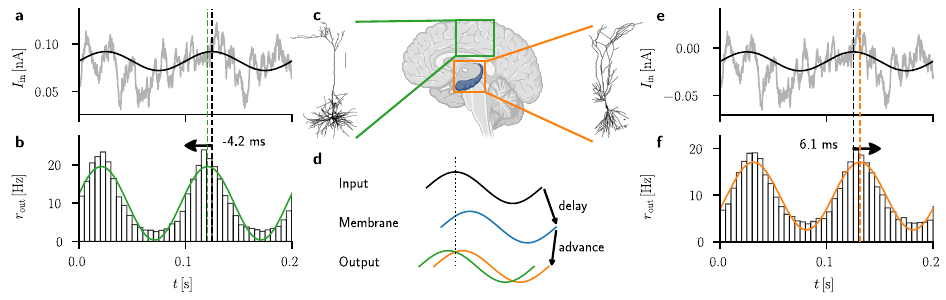}
    \caption{
    \textbf{Hodgkin-Huxley model parameters define whether neuronal firing is prospective or retrospective.}
    \textbf{(a, b)} The output firing rate $\rout$ of a neocortical neuron model \citep{mainen_model_1995} is phase-advanced by $4.2\,$ms with respect to the \SI{10}{Hz} amplitude modulation of a noisy input current $\Iin$ (leftward arrow).
    The vertical dashed lines indicate the time of a maximum of the input modulation (black) and of the corresponding sinusoidal fit of the output rates $\rout$ (green).
    Top: sample of $\Iin$ (grey). Bottom: averaged rate histogram.
    \textbf{(c)} Morphology of a neuron to which the parameters were fitted
    and their corresponding brain regions (green: neocortex \citep{mainen_model_1995}, orange: hippocampus  \citep{traub_neuronal_1991}).
    \textbf{(d)} While the membrane potential as a low-pass filtered input current is delayed with respect to the current, the output firing rate is advanced with respect to the voltage, overcompensating the delay (green, prospective with respect to the input) or not (orange, retrospective with respect to the input).
    \textbf{(e, f)} Analogously to (a, b), but for a hippocampal neuron \citep{brette_simulation_2007}, showing an average delay of $6.1\,$ms of the output rate with respect to the input modulation.
    }
    \label{fig:1_results}
\end{figure}

Cortical pyramidal neurons can fire with a phase-advanced in the range of several milliseconds with respect to their input current \citep{kondgen_dynamical_2008}. Even though the output rate is non-linearly related to the input, it is still possible to compare a sinusoidal fit of the output with the average sinusoidal modulation of the input, in order to detect a phase shift---indicating a temporal advance.
To test whether we can reproduce this finding with Hodgkin-Huxley-type neuron models, we simulated these neurons injecting an input current with the same characteristics as in \citet{kondgen_dynamical_2008}. The equations describing the Hodgkin-Huxley model are written down explicitly in Methods (Eq. (\ref{HHeqs}) et seq.).
\cref{fig:1_results}a-b shows an advanced firing response when using parameters fitted to cortical neurons (\cref{fig:1_results}c, green and Methods), in accordance to the experimental observation.
When we repeated the simulation with parameters fitted to hippocampal neurons (\cref{fig:1_results}c, orange) we observed that the firing response of the neuron is delayed (\cref{fig:1_results}e-f).
Thus, Hodgkin-Huxley neurons can show both advanced and delayed firing for input modulated with the same frequency.

Integrating the input current into the membrane voltage $\Vm$ naturally introduces a delay with respect to $\Iin$.
As the output rate $\rout$ depends on the membrane voltage, one would intuitively expect the output rate to also be delayed.
The fact that $\rout$ can be advanced with respect to $\Iin$ supports the hypothesis that individual neurons can implement prospective coding.
%For this, there must be some mechanism that compensates for the delay introduced by the membrane filtering.
To achieve a rate advance, the firing mechanism must over-compensate the delay introduced by the membrane filtering.
In general, the proportion between the delay (from input to membrane voltage) and the advance (from membrane voltage to output) defines whether a given neuron is in the prospective or retrospective regime (\cref{fig:1_results}d).

Here, we showed that the parameters of the model influence whether the output of a neuron is prospective or retrospective with respect to the input. In the following section, we will use this observation and the mechanistic interpretability of the Hodgkin-Huxley parameters to obtain a mechanistic explanation for prospective coding in individual neurons.

\subsection*{Parameters leading to prospective and retrospective coding}
To explore the parameters for which neurons can advance or delay their output with respect to their input, we sampled the parameter space for the Hodgkin-Huxley model using simulation-based inference (SBI, see \cite{tejero-cantero_sbi_2020}).
The most relevant parameters determining the temporal response properties turned out to be the leak conductance ($\gl$), the opening rate of the sodium ion channels ($\am$) involved in the action potential initiation, and the closing rate of the sodium ion channels ($\bh$, see Methods for a more detailed discussion). 

We verified that the three parameters $(\gl, \am, \bh)$ yield a diverse range of advances and delays through SBI. Parameters were drawn from distributions that produce the desired phase shifts (the so-called posterior distributions conditioned on the phase shifts). 
Running simulations with parameters sampled from these conditional posterior distributions, while using noisy input currents, yielded phase shifts centered around the conditioned time shifts (\cref{fig:2_posterior}j). The results confirm that the Hodgkin-Huxley model can produce various advances and delays with slight modifications of the three parameters.
The distribution of phase shifts is narrow for the advanced condition and becomes wider from instantaneous to delayed firing, indicating a stronger sensitivity to small fluctuations of the parameters the further we move away from the original neocortical parameters (which lead to advanced firing). 
A typical set of parameters for each posterior condition is shown in \cref{fig:2_posterior}a-i, together with an example of the noisy input trace.
The shift of the average firing rate $\rout$ (\cref{fig:2_posterior}d-f) reproduces the phase shift that we use as a condition for the posterior (\cref{fig:2_posterior}k).
Notably, the shift of the temporal response is visible even at the level of the output of a single neuron in a single trial. To check for this (\cref{fig:2_posterior}g-i), we compared the outputs of one neuron from each class to the same realization of input noise, and checked that indeed the spike trains from the advanced condition are prospective with respect to those of the instantaneous and delayed conditions. 
This indicates that prospective coding in these neurons cannot be regarded as a purely statistical effect.

\subsection*{Mechanism for prospective coding} 
\begin{figure}[H]
    \centering
    \includegraphics[width=1\linewidth]{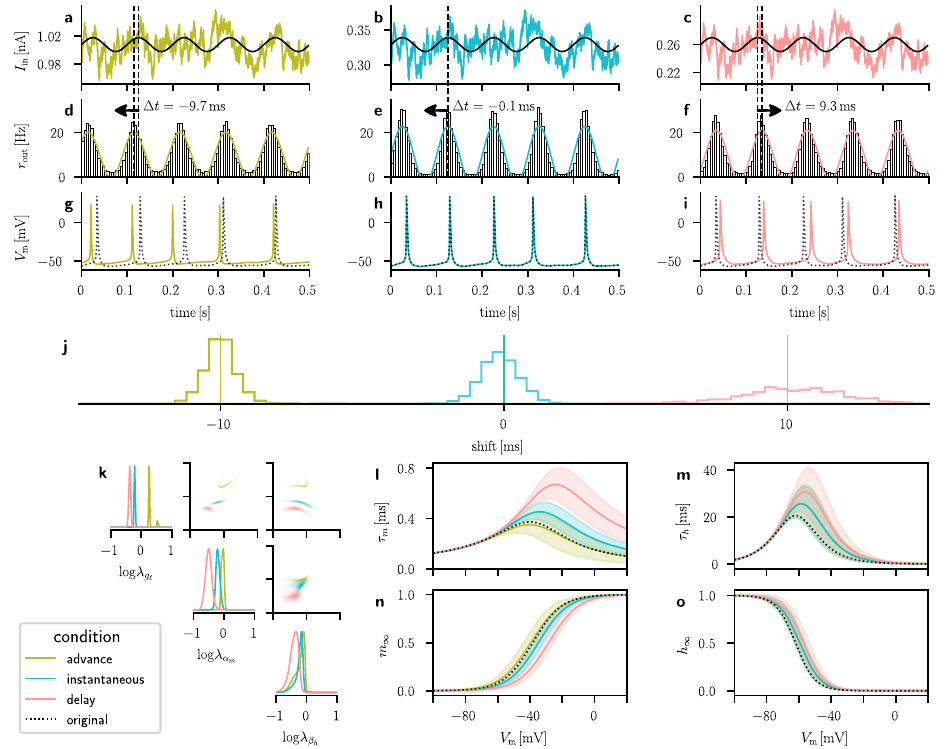}
    \caption{ 
    \textbf{
    Examples of prospective (green), instantaneous (blue) and retrospective (red) firing of a Hodgkin-Huxley neuron with different parameters.}
    {\bf (a-i)} Input currents with identical noise (a-c) that are sinusoidally modulated (black) produce an average firing rate that is advanced (d), instantaneous (e), or delayed (f) for different parameters $\am$, $\bh$ and $\gl$ of the sodium activation, inactivation and the leak conductance.
    (g-i): Voltage traces corresponding to the current shown above (dotted line identical to the instantaneous trace (h) for reference).
    {\bf (j)} The distribution of phase shifts, calculated from the input current to the output rate, as shown in (a-f), with parameters drawn from the posterior distribution conditioned on prospective ($-10\,$ms, green), instantaneous (\SI{0}{ms} blue) and retrospective ($10\,$ms, red) firing.
    {\bf (k)} The posterior densities of the parameters $\gl$, $\am$ and $\bh$, conditioned on a phase shift being advanced (green), instantaneous (blue) or delayed (red) as in j, while fixing the remaining parameters.
    The abscissa represents the logarithm of the factor $\lambda_\theta$ that scales the corresponding parameter $\theta$.
    The three diagonal plots show the conditional posterior densities of a single parameter, with the other two parameters fixed to the maximum of the joint density.
    The 2-dimensional plots show the densities of the parameter to the left and below, with the third parameter fixed.
    \textbf{(l -- o)} Time constants and steady states of the gating variables $m$ and $h$ as a function of the voltage, with mean and standard deviation obtained from sampling the joint parameter density conditioned on advanced (green), instantaneous (blue) and delayed (red) phase shifts.
    The black dotted line refers to the original parameters of the cortical neuron \citep{mainen_model_1995}.
    }
    \label{fig:2_posterior}
\end{figure}

To study the parameter dependencies of the response delays, we scaled the critical parameters $\am$, $\bn$ and $\gl$ by a corresponding scaling factor $\lambda$.  
The opening and closing rates $\alpha$ and $\beta$ of the gating variables $m$ and $h$ then define their time constants $\taum=1/(\lambda_{\am}\am+\bm)$ and $\tauh=1/(\ah+\lambda_{\bh}\bh)$, see \cref{fig:2_posterior}l-m, and steady states $\minf=\lambda_{\am}\am/(\lambda_{\am}\am+\bm)$ and $\hinf=\ah/(\ah+\lambda_{\bh}\bh)$, see \cref{fig:2_posterior}n-o.
It turns out that, to obtain an advanced response, both time constants need to be smaller compared to the instantaneous and delayed responses.
The smaller time constants lead to a faster activation and deactivation of the sodium channels.
Additionally, for the advanced response the steady states of the gating variables are shifted slightly to the left on the voltage axis, indicating an earlier opening of the $m$-gates and an earlier closing of the $h$-gates for a rising membrane potential---as during action potential initiation.
Because the activation and inactivation of the sodium channels are shifted to an earlier point and happen faster, the peak of the sodium conductance, and with it the peak of the action potential, occurs earlier.
The shift of the action potential to an earlier point in time then also advances the output rate of the neuron---a mechanism that is supported by the voltage traces shown in \cref{fig:2_posterior}g-i.
Additionally, the leak conductance $\gl$ influences the membrane time constant and a larger leak leads to faster input integration.

\subsection*{Advanced and delayed spike responses to stationary stochastic inputs}

We next study how robust the prospective firing of a neuron is for various kinds of inputs. So far we compared the temporal response of cortical neuron models to sinusoidally modulated noisy inputs.

\begin{figure}[H]
    \centering
    \includegraphics[width=.66\linewidth]{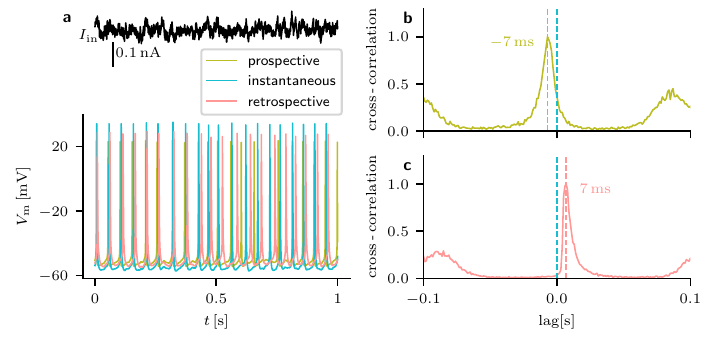}
    \caption{\textbf{Advanced and delayed spiking to the same stationary input current.}
    {\bf (a)} Voltage response to a stationary noisy current (black) for the 3 sets of parameters used in Figure \ref{fig:2_posterior}.
    The mean of the input is adapted to achieve the same average firing rates.
    {\bf (b, c)} Spike cross-correlogram for the prospective (b) and the retrospective (c) neuron parameters, with the spike times of the instantaneous parameter used as reference.
    The spikes of the prospective neuron are on average advanced by \SI{7}{ms}, and that of the retrospective neuron on average delayed by \SI{8}{ms}, compared to the spikes of the instantaneous neuron.
    The secondary peaks left and right correspond to the average inter-spike interval of \SI{0.1}{s}.
    The statistics is based on a total simulation length of \SI{1000}{s}.
    }
    \label{fig:3_crosscorr}
\end{figure}

Here, we investigate whether the advanced and delayed responses for the different parameter sets can be obtained without sinusoidal input modulation at all, just in the presence of the noisy input. In fact, an inspection of the spike timing shows that an advance or delay is independent of any modulation frequency (\cref{fig:3_crosscorr}). To quantify the temporal spike-shifts, we considered the cross-correlation function between the output of neurons with the different parameters to the same input current (\cref{fig:3_crosscorr}a).
The shift of the peak in the cross-correlation confirms the systematic earlier and later spiking of the neurons with the specific parameter sets as compared to a neuron showing an instantaneous spiking (\cref{fig:3_crosscorr}b-c).

\subsection*{The phase advance is robust to a wide range of inputs}
\label{sec:freq}
\begin{figure}[H]
    \centering
    \includegraphics[width=1\linewidth]{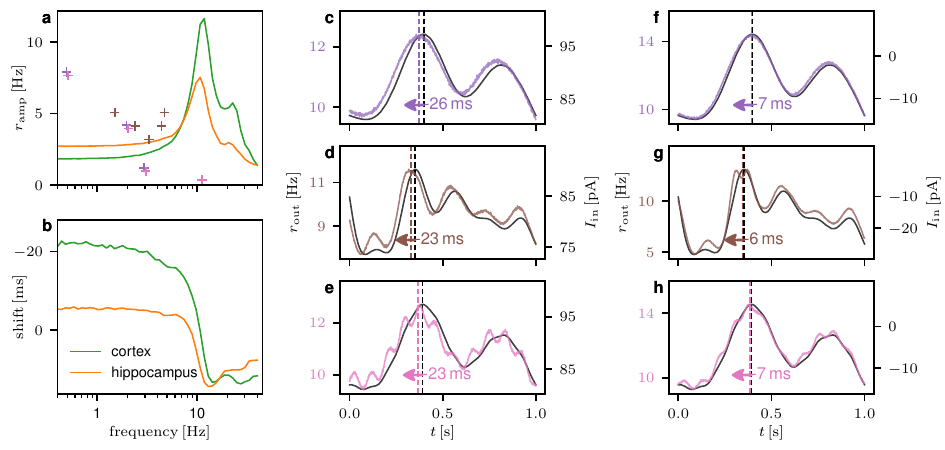}
    \caption{
    \textbf{The temporal advance is robust to a wide range of time-varying inputs.}
    \textbf{(a)} The amplitude response of a Hodgkin-Huxley neuron to sinusoidal input is frequency-dependent and shows a resonance frequency at \SI{11}{Hz} for the cortical parameters (green) and slightly less for the hippocampal parameters (orange).
    Colored crosses indicate the frequency and relative amplitude (a.u.) of the Fourier components composing the mean input currents in c-h.
    \textbf{(b)} Phase as a function of the frequency, showing an advance of $\rout$ for frequencies $\leq\SI{11}{Hz}$ (cortical parameters) and $<\SI{9}{Hz}$ for the (hippocampal parameters), and a delay for higher frequencies.
    \textbf{(c-e)} Averaged firing rates (colored) of the cortical neuron to the randomly modulated noisy input currents combining the Fourier components indicated in a (average current in black).
    Dashed lines with arrows indicate the phase advance computed from the cross-correlation between average current and rate.
    \textbf{(c)} With low frequency components, $\rout$ is largely advanced with respect to $\Iin$.
    \textbf{(d)} Including higher frequencies still shows an advance, although more skewed. \textbf{(e)} Adding an additional resonance frequency at \SI{11}{Hz} to the signal in (c), oscillations emerge in $\rout$.
    \textbf{(f-h)} Same as in c-e, but for the hippocampal neuron, showing only a modest advance.
    }
    \label{fig:4_sinusoidals}
\end{figure}

As biological neurons encounter complex input dynamics \emph{in vivo}, we further tested our neuron models for random-like inputs with a non-trivial frequency spectrum. 
For this, we first determined the amplitude and phase responses of our model neurons to sinusoidal inputs of various modulation frequencies (\cref{fig:4_sinusoidals}a, b).
As observed experimentally \citep{linaro_dynamical_2018}, the response amplitude shows a resonance frequency (\cref{fig:4_sinusoidals}a), and the phase advance can be more than \SI{20}{ms} for lower frequencies, while we see a delay above the resonance frequency (\cref{fig:4_sinusoidals}b).

It is interesting to study how this frequency dependency affects the response to random-like input currents. For this, we use as an input a superposition of sinusoidals with random frequencies, supplemented by noise (\cref{fig:4_sinusoidals}c-e).
We found that a signal consisting of a mixture of smaller frequencies leads to a clear phase advance of \SI{26}{ms} (\cref{fig:4_sinusoidals}c).
A signal with higher frequencies leads to an output rate $\rout$ that is slightly skewed but still shows advanced firing by \SI{23}{ms} with respect to the input current $\Iin$ (\cref{fig:4_sinusoidals}d).
To illustrate the effect of the resonance frequency, we added a component to the first signal with a frequency of \SI{11}{Hz} and a small amplitude that is barely visible in the input $\Iin$ but very prominent in the output rate $\rout$ and leads to a distorted response while the advance of \SI{23}{ms} is still dominant (\cref{fig:4_sinusoidals}e). Similar observations can be made for the hippocampal parameters, only that the advances (\SIrange{6}{7}{ms}, \cref{fig:4_sinusoidals}f-h) are smaller, as expected from the frequency response (\cref{fig:4_sinusoidals}b).

The resonance frequency for both parameter sets is close to the average output rate of the neurons that we selected to be \SI{10}{Hz} by adapting the mean input current.
We generally expect a neuron to resonate close to the average firing rate as the average inter-spike interval matches the period of the input modulation.
Accordingly, the resonance shifts to higher frequencies if we increase the average input current, leading to a wider frequency spectrum (smaller than the resonance frequency) for which prospective firing can be observed (\cref{fig:A4_frequency_response_input_dependency}).
In every case, the interesting regime for prospectivity occurs close to this resonant frequency. Whereas low frequency consistently elicits an advanced response, even for the hippocampal neuronal parameters (\cref{fig:4_sinusoidals}b), and high frequencies are consistently retrospective, the shape of the transition between these two modes is what characterizes the complex temporal responses. In the case of cortical and hippocampal parameters, it is only around the resonant frequency that we find a clear separation of prospective and retrospective responses. 

\subsection*{Adaptation implies prospectivity with respect to the voltage}
Having described the mechanism behind the prospective nature of Hodgkin-Huxley neurons, we next introduce an approximation in terms of a simplified rate model.
This allows us to show how prospective coding can originate from the behavior of the $h$ variable of the Hodgkin-Huxley model, which resembles an adaptation mechanism. Besides simplicity, this resemblance has the advantage that it easily generalizes to other rate models.
Because the time course, in particular the positive deviations, of the sodium conductance reflects the occurrence of action potentials, they constitute a good approximation of the time course of the output rate of a neuron (\cref{fig:A5_r_out_dependency}).
Thus, to simplify the Hodgkin-Huxley model we consider the output rate as a function of the sodium conductance, $\rhh = \varphi(\gNa)$.
We approximate $\gNa$ for small frequencies and small time constants (i.e.~small $\tauh$) in terms of the steady-state sodium conductance $\gNa^\infty$ and its temporal derivative,
\begin{equation}
    \begin{split}
        \rhh
        = \varphi(\gNa)
        &\approx \varphi(\gNa^\infty(v) + \trhh(v)\,\dot{g}_{\rm Na}^\infty(v)) \\
        &\approx \varphi(g_{\rm Na}^\infty(v + \trhh(v)\,\dot v)) \\
        &= \varphi_\mathrm{HH}(v + \trhh(v)\,\dot v)
        = \rhhpros \;.
        \label{eq:rdyn}
    \end{split}
\end{equation}
Here, the approximation in the second line can be seen as a first-order Taylor expansion, which allows us to shift the temporal dynamics into the argument of $\gNa^\infty$.
The time constant $\trhh(v)$ of the sodium conductance is positive in the voltage regime of interest and thus describes a prospective mechanism (Methods).
While $\rhh$ refers to the full dynamics of $\gNa=m^3h$ with no approximations, $\rhhpros$ is our approximated model.
Note that it is based on the prospective voltage $v+\trhh(v)\,\dot v$.
Because $\rhhpros$ is a function of the prospective voltage, its time course is advanced with respect to the original voltage and hence, our model predicts that also $\rhh$ is advanced.
Up to small deviations around higher rates, simulations confirm that $\rhhpros$ approximates $\rhh$ well (\cref{fig:5_theory}a).
Crucially, both rates show a similar advance compared to the 'instantaneous rate' $\rhhinf = \varphi(g_{\rm Na}^\infty (v) ) = \varphi_\mathrm{HH}(v)$, which is a function of the voltage $v$ only.
The advance of $\rhhpros$ is explained by the $\dot v$ contribution, inducing an oscillation of $\rhh$ and $\rhhpros$ around the steady state $\rhhinf$ (\cref{fig:5_theory}b).

Finally, we verify our prediction that the approximations in our model lead to divergences between $\rhhpros$ and $\rhh$ if either $\tauh$ or $\dot v$ is large by increasing the frequency of a sinusoidal input and thus increasing $\dot v$.
The temporal shift based on $\rhhpros$ approximates the shift estimated with $\rhh$ well for small frequencies (\cref{fig:5_theory}c) and diverges for increasing frequencies until they both reach zero again for high frequencies, qualitatively reproducing the temporal advance of the spiking Hodgkin-Huxley model in \cref{fig:4_sinusoidals}b.

\cref{eq:rdyn} shows that the firing rate is advanced with respect to the voltage with a time advance in the order of $\trhh(v)$. Yet, with respect to the input current $I$, the rate can be both prospective or retrospective.
This is because the voltage lags behind the input by the membrane time constant $\tau_{\rm mem}$.
The proportion between $\tau_{\rm mem}$ and $\trhh$ determines whether the output is prospective or retrospective with respect to the input, as described in \cref{fig:1_results}d.

\subsection*{Slow threshold adaptation implies long prospective coding}
Prospective coding, in general, can result from adaptation-like processes on different time scales. We illustrate this observation with the example of threshold adaptation that operates on a time scale 10 times longer than the adaptation in the Hodgkin-Huxley model, generating correspondingly greater advances in the output response.

To give an intuition of why the adaptation of the firing threshold $\vartheta$ leads to prospective firing, we write the firing rate as $r = \psi(v - \vartheta) = \psi(v_\vartheta)$, where $v_\vartheta$ is the voltage relative to the threshold.
The threshold dynamics is defined such that $\vartheta$ slowly increases with a rising voltage, and thus counteracts the rising voltage by a delayed threshold increase.
Similar as before for the Hodgkin-Huxley dynamics, we approximate the threshold function $v_\vartheta$ in terms of the steady state $v_\vartheta^\infty(v)$ and its temporal derivative to obtain the output rate as a function of the prospective voltage,
\begin{equation}
    \begin{split}
        \rthr
        = \psi(v_\vartheta)
        &\approx \psi(v_\vartheta^\infty(v) + \trthr(v)\,\dot{v}_\vartheta^\infty(v)) \\
        &\approx \psi( v_\vartheta^\infty(v + \trthr(v)\,\dot v) ) \\
        &= \psi_\vartheta(v + \trthr(v)\,\dot v)
        = \rthrpros \;.
    \end{split}
 \label{eq:thr_pros}
\end{equation}
As before, the approximation in the second line uses a first-order Taylor expansion to move the temporal dynamics into the argument of $v_\vartheta^\infty$.
The adaptation time constant $\trthr(v)$ is strictly positive and determines the prospective rate $\rthrpros$, which is a function of the prospective voltage $v + \trthr(v)\,\dot v$.

\begin{figure}[H]
    \centering
    \includegraphics{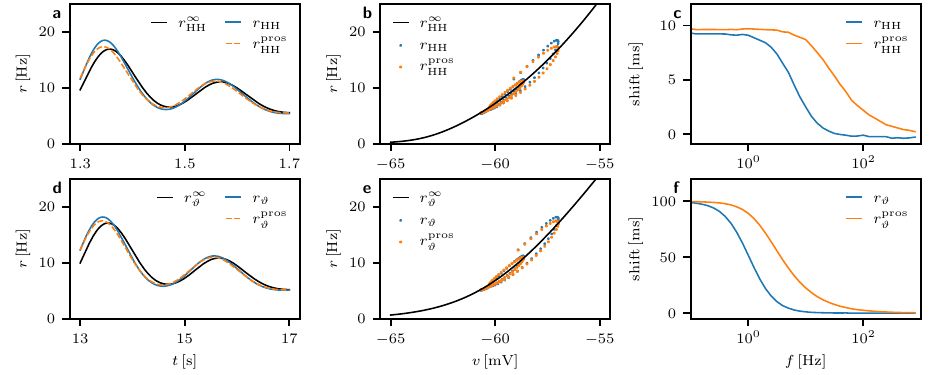}
    \caption{
    {\bf A fast spike- and slow threshold-adaption lead to temporal advances on different time scales.}
    {\bf (a)} The output rate of a neuron as a function of the sodium conductance $\rhh = \varphi(\gNa)$ (blue) is prospective with respect to the instantaneous rate $
    \rhhinf(v) = \varphi(g_{\rm Na}^\infty(v))$ (black).
    The approximated rate $\rhhpros$ matches $\rhh$ well and shows a similar advance (dashed orange).
    {\bf (b)} Firing rates $\rhh$ and $\rhhpros$, sampled every $50\,$ms (blue dots), turn clockwise around the steady-state voltage-to-rate transfer function ($\rhhinf(v)$, black).
    {\bf (c)} Temporal advance of $\rhh$ (blue) and the approximation $\rhhpros$  (orange) with respect to the voltage for different frequencies.
    The approximation qualitatively reproduces the advance that decays with increasing frequencies.
    {\bf (d-f)} Corresponding rate functions for the threshold adaptation model, with $\rthr = \psi(v -\vartheta)$ (blue) and a dynamic threshold $\vartheta$ being the original model, $\rthrpros = \psi_\vartheta(v + \trthr \dot v)$ (dashed orange) its approximation, and the instantaneous rate $\rthrinf(v)=\psi(v_\vartheta^\infty(v))$ (black), in analogy to the sodium inactivation in (a-c).
    The longer time scale (d, as compared to a) reflects the slower modulation of the input in the threshold adaptation model, resulting in a 10 times larger temporal advance (f, $100\,$ms) as compared to the sodium inactivation model (c, $10\,$ms).
    }
    \label{fig:5_theory}
\end{figure}

Because the rate $\rthrpros$ is a function of the prospective voltage, it is advanced on the order of $\trthr(v)$ with respect to the instantaneous steady state of the rate $r_\vartheta^\infty = \psi( v_\vartheta^\infty (v) )= \psi_\vartheta^\infty(v)$.
Following \cref{eq:thr_pros}, the adaptation of the threshold causes $\rthr$ to be advanced similarly (\cref{fig:5_theory}d-f).

To provide a rough intuition for the time advance, consider any dynamic negative adaptation, that is an adaptation that counteracts the membrane voltage with a delay.
The contribution of the delayed adaptation to the output response leads to dependency on the temporal derivative of the voltage.
This dependency, which is positive for negative adaptation (Methods), causes the output response to be more sensitive to changes in the voltage.
Whenever the voltage rises, the prospective voltage rises faster and it descends faster for a descending voltage, effectively shifting the maximum to an earlier point in time. An alternative calculation of the full spectrum of adaptation-induced phase shifts can be found in \citep{ellenberger_backpropagation_2024}.

It is worth emphasizing that adaptation-like processes, be it the sodium adaptation in \cref{eq:rdyn} or the threshold adaptation in \cref{eq:thr_pros}, are formally similar. Dendritic adaptation in terms of a negative dendritic $I_h$ current that counteracts the dendritic depolarization \citep{kole_single_2006} is yet another adaptation process that formalizes in the same way, namely as a rate $r= \psi(v+I_h)$. Such a dendritic current is advancing a dendritic voltage response while propagating to the soma \citep{ulrich_dendritic_2002}, see Methods.

\section*{Discussion}
In this work, we elucidated cellular mechanisms which could explain prospective firing in cortical neurons.
Specifically, (i) we showed that the Hodgkin-Huxley model can reproduce experimental \emph{in vitro} results involving prospective and retrospective firing in cortical and hippocampal neurons.
(ii) By studying the parameter space of Hodgkin-Huxley model neurons around cortical parameters, we characterized the fast activation and deactivation of sodium channels as a mechanism behind prospectivity at timescales of tens of milliseconds.
(iii) Prospectivity and retrospectivity depend on the modulation frequency of the input. For sinusoidal-like input, lower frequencies elicit prospective responses of the rate with respect to the input current, whereas higher frequencies elicit retrospective responses.
(iv) A simplified model for prospectivity based on a look-ahead voltage is compatible with these results and generalizes them to larger timescales.
(v) Our model shows that adaptation translates into a temporal advance in response to slowly changing current inputs. Negatively coupled adaptation-like variables trigger an anti-inertial behavior on the system, accelerating changes both for increasing and for decreasing voltages.

Various experimental findings show prospective rates on different time scales. On time scales of several milliseconds, 
recordings of firing rates in response to noisy periodic input currents showed frequency-dependent phase advances  \citep{kondgen_dynamical_2008, lundstrom_fractional_2008, ostojic_neuronal_2015,linaro_dynamical_2018}.
On longer timescales, \citet{fuhrmann_spike_2002, pozzorini_temporal_2013} explicitly showed how spike frequency adaptation also leads to a prospective output at the level of individual neurons.
Other types of adaptation, for instance caused by dendritic currents, may also contribute to prospective coding \citep{senn_neuronal_2024}. Inspired by the diversity of adaptation mechanisms, we suggest a mechanism-agnostic description of prospectivity at the neuronal level that takes into account the rate of change of the membrane voltage. Our model highlights how different prospective mechanisms allow for prospectivity on different timescales, capturing all types of fast and slow adaptation through the formalization of the prospective voltage. 

In contrast to the observation that time shifts depend on the frequency of the input and saturates for low frequencies, several previous studies---which interpreted prospective firing as a result of fractional differentiation and power-law dynamics \cite{pozzorini_temporal_2013, lundstrom_fractional_2008}---yield a constant phase advance \cite{lundstrom_fractional_2008}. A constant phase advance would lead, in contrast to our results, to an increasing time shift for lower frequencies. Moreover, a constant phase advance is not in accordance with experimental results \cite{kondgen_dynamical_2008}. A possible explanation for these discrepancies is that these previous works only considered comparatively low frequencies.

Prospectivity may be important for several cognitive functions. Prospective firing is dependent on input frequency modulation, and in the brain, different frequencies are dominant depending on state and function.  For example, low frequencies below $10\,$Hz are implicated in recalling and attention tasks \cite{castelhano_intracranial_2022}, and for these frequencies we show robust prospectivity. The mechanism may also be exploited by neural networks in artificial intelligence to solve temporal processing tasks. Recent theoretical work exploited prospective firing of neurons in larger networks to overcome delays, to perform temporal computation, and to build bio-inspired hardware \citep{senn_neuronal_2024, haider_latent_2021, ellenberger_backpropagation_2024}.

In terms of experimental predictions, our model posits that the mechanism underlying the type of prospectivity observed by \citet{kondgen_dynamical_2008} is tightly linked to the opening and closing rate of sodium channels. Because the opening and closing rates of ion channels increase with temperature \citep{ulbricht_sodium_2005} (and can be modulated by drugs \citep{wisedchaisri_druggability_2022}), our model predicts that prospectivity decreases with temperature, which could be tested in a simple experiment. 
More generally, the model predicts that any type of adaptation that modulates the membrane potential will induce prospective firing on the corresponding time scale.

\section*{Methods}
\subsection*{Hodgkin-Huxley model}
\label{sec:HH}
The Hodgkin-Huxley model \citep{hodgkin_quantitative_1952} remains one of the most important representations of action potential generation in neurons.
It incorporates voltage-gated sodium and potassium channels, which are modeled by four gating particles each.
The variables $m$, $h$ and $n$ each describe the fraction of corresponding gating particles in the open configuration.
The total conductance of a channel is usually modeled by the maximal conductance (e.g., $\bar{g}_{\rm Na}$) times the fraction of open channels which is defined by the fraction of gating particles in the open conformation (e.g., $\gNa=\bar{g}_{\rm Na}m^3h$).

During the initial phase of the action potential and the rising phase of the membrane potential $\Vm$, the fast opening of the sodium channels through the $m$ gates leads to an increased influx of sodium ions which causes a steep increase in $\Vm$ (depolarization).
The subsequent closing of the $h$ gates leads to a closing of the sodium channels and stops the influx of sodium ion.
Meanwhile, the potassium channels open through the slow opening $n$ gating particles and the efflux of potassium ions leads to a repolarization and hyperpolarization of $\Vm$ in the late phase of the action potential.

The Hodgkin-Huxley model is described by the dynamics of the membrane potential $\Vm$:

\begin{equation}\Cm \dot{V}_{\rm m} = \gl\,(\El - \Vm) + \bar{g}_{\rm Na}m^3h\,(E_{Na} - \Vm) + \bar{g}_{\rm K}n^4\,(E_K - \Vm) + \Iin,
\label{HHeqs}
\end{equation}
where $\Cm$ is the membrane capacitance, $\gl$ and $\El$ are the leak conductance and reversal potential, $\bar{g}_{Na}$ and $\bar{g}_{K}$ the maximum conductances of the sodium and potassium channels, $E_{Na}$ and $E_K$ their reversal potentials and $\Iin$ is the input.

The dynamics of the voltage dependent gating variables can be described by

\begin{align}
    x = x_\infty(\Vm) - \tau_x(\Vm)\,\dot{x},
\end{align}

for $x\in\{m,n,h\}$, with the steady state $x_\infty$ and the time constant $\tau_x$ both voltage dependent.
The variable $x$ describes the fraction of gating particles in the corresponding channel type that are in the open state. An alternative equivalent formulation that might be more intuitive is

\begin{align}
    \dot{x} = \alpha_x(\Vm)\,(1-x) - \beta_x(\Vm)\,x,
\end{align}

which uses the voltage dependent opening rate $\alpha_x(\Vm)$ and closing rate $\beta_x(\Vm)$.
The first term corresponds to the fraction of particles in the closed state $(1 - x)$ that opens with rate $\alpha_x(\Vm)$.
Similarly, the second term corresponds to the fraction of particles in the open state $x$ that close with rate $\beta_x(\Vm)$.
The steady state of the gating variable is then $x_\infty(\Vm)=\alpha_x(\Vm)/(\alpha_x(\Vm)+\beta_x(\Vm))$ and the time constant $\tau_x(\Vm)=1/(\alpha_x(\Vm)+\beta_x(\Vm))$.

The Hodgkin-Huxley model can be used to describe different types of neurons in different parts of the brain with the parameters fitted to experimental data of patch clamp recordings.
In this study, we use two sets of parameters fitted to cortical and hippocampal pyramidal neurons.
The cortical parameters, were taken from \cite{gerstner_neuronal_2014}, with $m$ and $h$ fitted by \cite{mainen_model_1995}, on experiments reported by \cite{huguenard_developmental_1988}, and $h$ fitted by Richard Naud on experiments reported by \cite{hamill_patch-clamp_1991}.
For the hippocampal parameters we relied on an adaptation by \cite{brette_simulation_2007}, of the model by \cite{traub_neuronal_1991}. The detailed parameters are shown in the SI.

\subsection*{Simulations}
\label{sec:model}
For our numerical experiments we simulate the Hodgkin-Huxley model with different parameters.
We rely on the \texttt{brian2} package \citep{stimberg_brian_2019}, for the simulation of differential equations in the context of neuronal experiments.
We performed simulations similar to the experiments in \cite{kondgen_dynamical_2008}, injecting a sinusoidal input current with Ornstein-Uhlenbeck noise into the neurons membrane:

\begin{align}
    \Iin = I_0 + I_1 \sin(2\pi f t) + I_\textrm{noise},
\end{align}
where

\begin{align}
    \dfrac{I_{\rm noise}(t + \dt) - I_{\rm noise}(t)}{\dt} = - \dfrac{1}{\tau_\textrm{noise}}I_{\rm noise}(t) + \sigma\,\sqrt{\dfrac{2}{{\rm d}t\,\tau_{\rm noise}}}\mathcal{N}(0,1)
\end{align}

is an Ornstein-Uhlenbeck process with time constant $\tau_\textrm{noise}$, variance $\sigma^2$ and $\mathcal{N}(0,1)$ is the normal distribution with zero mean and unit variance.
Thus, the input current is sinusoidal with mean $I_0$, signal amplitude $I_1$, noise amplitude $\sigma$ and $\tau_\textrm{noise}=\SI{10}{ms}$.
During the simulation, the refractory period, a short time interval during which the input current is not integrated into the neuron's membrane potential, is defined by $m>0.5$.
During experiments we keep $I_1=\SI{0.01}{nA}$ and $\sigma=\SI{0.02}{nA}$ fixed.
If not otherwise stated, we adapt the mean input current $I_0$ such that the average $\rout$ is \SI{10}{Hz}.
The simulation length is usually \SI{1.5}{s}, of which the first second is used for initialization and then discarded.

The response of the neuron to the input current is examined by analysing the spike times of the membrane potential.
We do not consider sub-threshold stimulation, for which the input would be an ideal reference, but rely on the spikes of the neuron as we investigate the action potential generation as a mechanism that leads to prospective output rates.
We compute the number of spikes after each simulation and define the spike time using the maxima of the action potential.
To estimate the output firing rate $\rout$ of a neuron, the peristimulus time histogram (PSTH) is computed over repeated trials, usually with a resolution of 100 bins if not stated otherwise.
For each trial, the simulation is repeated with identical parameters but different noise.
For simplicity, we run one simulation with multiple identical neurons that receive the same signal but different noise and that have no interconnections.

For the random-like signals in \cref{fig:4_sinusoidals}, we use a combination of different frequencies and amplitudes.
For the first signal, that should demonstrate the advance for slow frequencies, we chose frequencies $f_1=0.5$, $f_2=2$ and $f_3=3$ $\SI{}{Hz}$ with amplitudes $a_1=-1$, $a_2=-0.5$ and $a_3=0.1$.
The second signal is more interesting and contains a wider range of frequencies: $f_1=1.5$, $f_2=2.4$, $f_3=3.3$, $f_4=4.5$ and $f_5=4.7$ $\SI{}{Hz}$ with amplitudes $a_1=0.5$, $a_2=0.4$, $a_3=0.3$, $a_4=0.4$ and $a_5=0.5$.
The third signal is the same as the first but additionally contains a frequency close to the resonance with $f_4=11$ with a low amplitude of $a_4=0.02$ that leads to an over-expressed response in the output rate because of the resonance.
The final modulation of the input rate is set to be zero mean and unit variance over time:
\begin{align}
    s^*(t) &= \sum_i a_i\,\sin(2\,\pi\,f_i\,t) \\
    s(t) &= \dfrac{(s^*(t) - \mathbb{E}[s^*])}{\mathrm{Var}[s^*]}.
\end{align}
With the expected value $\mathbb{E}$ and variance $\mathrm{Var}$ computed over time. The input is then in these cases $\Iin = I_0 + I_1 \, s + I_\textrm{noise}$.
To get a precise estimate of the instantaneous firing rate $\rout$, we simulate 10,000,000 trials and compute the PSTH with 1,000 bins.

For the theoretic adaptation models shown in \cref{fig:5_theory}, we use an excerpt of the second signal described above and shown in \cref{fig:4_sinusoidals}d and e.
For the Hodgkin-Huxley model, we use the cortical parameters for the variables $\minf$, $\hinf$ and $\tauh$ as well as to simulate $m$ and $h$.
We obtain the voltage $v$ via the leaky integrator dynamics $\Cm \dot v = \gl\,(\El - v) + \Iin$, where we use again the Hodgkin-Huxley cortex parameters for $\Cm$, $\gl$ and $\El$. The input is $\Iin = I_0 + I_1\,s$, with the signal $s$ as described above, the mean input $I_0$ is set such that the mean of $v$ is \SI{-60}{mV} and $I_1$ set such that the amplitude of $v$ is \SI{2.5}{mV} for a signal with amplitude 1.
To obtain the rate $\rhh$ from the sodium conductance $\gNa$, we chose a softplus activation function of the form $\varphi(x) = \dfrac{1}{\beta}\,\log(1 + \exp(\beta\,(a\,x+b))$.
Here, $\beta=1$ defines the sharpness of the transition from zero to linear, while we use $a=\SI{5e4}{}$ and $b=-5$ to shift the sodium conductance such that we get reasonable output rates.

For the threshold adaptation, we have more freedom to chose parameters and variables.
Yet, to consider dynamics that are to some extent biologically plausible, we rely on adaptations of the Hodgkin-Huxley variables.
For the adaptation we consider, the threshold should increase when the voltage increases.
This is the case for $\minf$, which is a sigmoidal and has a strictly positive derivative.
We write the steady state of the threshold as

\begin{align}
    \vartheta_\infty (v) = \vartheta_0 + \dfrac{\Delta_\vartheta}{1 + \exp(-2\,\dfrac{v - E_\vartheta}{\Delta_\vartheta})},
\end{align}
with the minimum $\vartheta_0=\SI{-55}{mV}$, the slope factor $\Delta_\vartheta=\SI{30}{mV}$ and the reversal point $E_\vartheta=\SI{-60}{mV}$.
As time constant $\tau_\vartheta$, we scale $\taum$ such that $\trthr$ in \cref{eq:rdyn} is \SI{100}{ms} for the mean voltage $v_0=\SI{-60}{mV}$.
Again, we consider a softplus function to map the distance to the threshold $v_\vartheta$ to the output rate. The parameters are $\beta=0.1$, $a=\SI{1e4}{}$ and $b=125$.

\subsection*{Simulation-based inference}
To understand how the parameters of the Hodgkin-Huxley model affect the phase response of a neuron, we use the simulation-based inference (SBI) package \citep{tejero-cantero_sbi_2020}.
We define a prior distribution $p(\theta)$ for each parameter, a Gaussian distribution centered around the respective fit to cortical neurons (\cref{fig:1_results}).
Next, we sample from the priors and run simulations from which we get a phase shift $\Delta t$ for each set of parameters, which allows us to sample the likelihood $p(\Delta t|\theta)$.
Given the priors and the simulations, we use SBI to estimate a posterior distribution $p(\theta|\Delta t)$ using neural density estimation.
To simplify the parametrization, we use scaling factors $\lambda$ as SBI parameters $\theta$.
While we could directly sample the parameters of the model (e.g. $\gl$), the equations for the gating variables (e.g. $\am$) have multiple parameters themselves and using a single scaling factor here reduces the complexity of the parameter space significantly.
To follow a consistent approach, we therefore use scaling factors for all parameters and equations.
Explicitly, for the scaling factors, we sample exponents from a normal distribution $\theta_i\sim\mathcal{N}(0, 0.1)$ such that we get $\lambda_i = 10^{\theta_i}$, to account for the multiplicative nature of scaling factors.
To get an intuition of how the scaling factors change the gating variables see \cref{fig:A1_scaling_factors}.

For each sampled set of parameters, we run 10,000 repeated trials to obtain an accurate estimate of $\rout$ and $\Delta t$, defined as the temporal shift between $\Iin$ and $\rout$.
The mean input current $I_0$ is adapted so that the average $\rout$ is \SI{10}{Hz}.
For this, we inject a constant current into the neuron that we progressively increase until the inter-spike interval is \SI{100}{ms}, matching the average output rate of \SI{10}{Hz} for sinusoidally modulated input.
In the process, we reject sets of parameters that do not lead to action potentials.
Additionally, we reject sets of parameters for which we don't find an input current that leads to an average firing rate between \SIrange{5}{15}{Hz} or if the resting potential for the parameters is above \SI{-30}{mV} while increasing $I_0$, which is usually a sign of a broken spike generation mechanism because of a bad parameter combination.
For algorithmic stability, we restrict $I_0$ to \SIrange{-10}{10}{nA}, which is a range large enough to cover all valid parameter sets.
To make parameter sampling more efficient, and avoid invalid parameters, we use a built-on feature of the SBI package to update the prior every 1,000 simulations.

We start with a set of 9 parameters, $\theta$, relevant for the spiking mechanism of the Hodgkin-Huxley model.
These are the leakage conductance $\gl$, the sodium and potassium conductance $\gNabar$ and $\gKbar$ and the opening and closing variables of the ion channel gating particles $\am$, $\bm$, $\ah$, $\bh$, $\alpha_n$ and $\beta_n$ (see \cref{fig:A2_conditional_posterior_9d}).
To discern which among these parameters are important for distinguishing prospective and retrospective firing, we compare the posterior distribution $p(\theta | \Delta t)$, conditioned on temporal phase shifts $\Delta t$.
The posterior distributions of some of the parameters ($\gKbar$, $\an$ and $\bn$) cover most of the sampling space, indicating that they are not relevant as changing them has little influence on the outcome.
The parameters that influence the temporal response should have narrow posterior distributions that do not overlap for the different observation.
The parameters we determine to be relevant are $\gl$, $\am$ and $\bh$.
Here, we omit $\bm$ as one of the relevant parameters, as it shows a linear dependency with $\am$ indicating that the two parameters are correlated and have the same informative value.

Continuing with the reduced set of three parameters, we sample 10,000 sets of parameters for the SBI analysis from which 982 sets were discarded, in most cases before the first restriction of the prior. Valid simulations are then used to estimate the posterior distribution for the three parameters.

The posterior distribution can be conditioned on any observation or phase shift $\Delta t_{\rm o}$, such that we can sample parameters $\theta_{\rm o}\sim p(\theta|\Delta t_{\rm o})$.
Running simulations of Hodgkin-Huxley neurons with the parameters $\theta_{\rm o}$ should yield the phase shift $\Delta t_{\rm o}$ that was used to condition the posterior.
For this posterior check, we sample 1000 sets of parameters for each condition and show that the model we use indeed reproduce the phase shifts we conditioned on (\cref{fig:2_posterior}j).

For the conditional posterior plots, either one or two parameters are varied, while all other parameters are fixed, resulting in either a 1-dimensional or 2-dimensional distribution (\cref{fig:2_posterior}k and \cref{fig:A2_conditional_posterior_9d}).
Here, we use a maximum likelihood approach to set the fixed parameters to the maxima of the conditional posterior distribution.
We show in total three different conditional posteriors, each with a different color and the transparency reflecting the probability.

For the individual simulations shown in \cref{fig:2_posterior}a-i, for each temporal shift we chose one set of parameter that is typical, i.e., a random representative of the posterior.
The parameters are $\lambda_{\gl}=1.6261$, $\lambda_{\am}=0.5310$ and $\lambda_{\bh}=0.4266$ for the prospective condition (green), $\lambda_{\gl}=0.7057$, $\lambda_{\am}=0.4572$ and $\lambda_{\bh}=0.3331$ for the instantaneous condition (blue) and $\lambda_{\gl}=0.4295$, $\lambda_{\am}=0.3344$ and $\lambda_{\bh}=0.3400$ for the retrospective condition (red).

In the case of the steady states and time constants of the gating variables shown in \cref{fig:2_posterior}l-o, we plot the mean over parameters and their standard deviation within each posterior distribution.
We consider the same parameters that where used to obtain the distributions of phase shifts shown in \cref{fig:2_posterior}j.

\subsection*{Signal analysis}
Following \cite{kondgen_dynamical_2008}, the response of a neuron $\rout$ can be described in the linear regime by

\begin{align}
    r(t) \cong r_0 + r_1(f)\,\sin(2\pi ft + \Phi(f)),
\end{align}

where the amplitude response $r_1(f)$ and the phase response $\Phi(f)$ are frequency dependent.

To estimate the amplitude and frequency response of a neuron to sinusoidal input, we fit a sinusoidal function to the output rate using least squared errors and the \texttt{scipy} package \citep{virtanen_scipy_2020}.
In \cref{fig:1_results}a and b, the shift between the sinusoidal input and the sinusoidal fit of the output rate is shown.
Even though the response of a neuron can be non-linear, we expect the temporal shift to be well estimated by a sinusoidal fit, as maxima of the fit will coincide with the maxima of the output rate for a least square fit.

For the cross-correlogram plot in \cref{fig:3_crosscorr}b and c, we compute the cross-correlation between two spike trains with the \texttt{scipy} package \citep{virtanen_scipy_2020} via
\begin{equation}
    c(k) = (x * y)(k - N + 1) = \sum_{l=0}^{N - 1} x_l\,y^*_{l-k+N-1}.
\end{equation}
Because we use a discrete cross-correlation, we discretize the spike times into bins of ${\rm d}t=\SI{e-4}{}$ with each element of the array being equal to one when a spike occurred and zero otherwise.
When the signal is not sinusoidal, we use the maximum of the cross-correlation as an estimate of the temporal shift between two signals as in \cref{fig:4_sinusoidals}.

\subsection*{Theoretical models} 
For the approximation of the Hodgkin-Huxley dynamics, we assume the output rate to be a function of the sodium conductance $r = \varphi(\gNa) = \varphi(m^3h)$, with sodium activation variable $m$ and inactivation variable $h$.
Since the activation is fast, we approximate $m$ to be an instantaneous function of the voltage, $m \approx \minf(v)$.
We make a marginally less crude approximation of the dynamic inactivation variable $h$ by

\begin{align}
    h = \hinf(v) - \tauh(v)\,\dot{h} \approx \hinf(v) - \tauh(v)\,\dot{h}_\infty(v),
\end{align}
where we replaced $\dot h \approx \dot{h}_\infty = h'_\infty\,\dot v$.
This leads to the approximation of the sodium conductance:
\begin{equation}
        \gNa = m^3 h \approx \minf^3(v)\,(\hinf(v) - \tau_h(v) \, h'_\infty(v)\,\dot v) = \gNa^\infty(v) + \trhh(v)\,\gNa^{\infty\prime}(v)\,\dot v = \gNa^\infty(v) + \trhh(v)\,\dot{g}_{\rm Na}^\infty\
         % = g_\mathrm{N}(v, \dot v),
    \label{eq:gNa}
\end{equation}
where $\gNa^\infty(v)=\minf^3(v)\,\hinf(v)$ is the steady state and $\trhh(v)=-\dfrac{\tauh(v)\,\minf^3(v)\,\hinf'(v)}{\gNa^{\infty\prime}(v)}$ the effective time constant of the sodium conductance, both functions of $v$.

To express the output rate of a neuron in terms of the prospective voltage, we move the temporal dynamics into the sodium steady state and do a first order Taylor expansion for small voltage fluctuations $\gNa^\infty(v + \trhh(v)\,\dot v) \approx \gNa^\infty(v) + \trhh(v)\,\gNa^\infty(v)'\,\dot v$, which is the right hand side of \cref{eq:gNa}.
With this, we rewrite the estimate of the output rate in our model as
\begin{equation}
        \rhh  = \varphi(\gNa) \approx \varphi(g_{\rm Na}^\infty(v + \trhh\,\dot v)) = \varphi_\mathrm{HH}(v + \trhh \dot v) \, = \rhhpros.
        \label{eq:rdyn_methods}
\end{equation}
Here, $\rhh$ refers to the full dynamics of $\gNa=m^3h$ with no approximations and $\rhhpros$ is our approximated model depending on the prospective voltage $v+\trhh\dot v$.
We compare $\rhh$ and $\rhhpros$ to the instantaneous rate $\rhhinf(v) = \varphi(g_{\rm Na}^\infty (v) ) = \varphi_\mathrm{HH}(v)$ because the nonlinear activation function shown in \cref{fig:5_theory}b leads to skewed rates and makes a direct comparison to $v$ impractical.
While we do not consider the effect of the sodium conductance on the membrane voltage, the feedback into the leaky integrator dynamics would advance the membrane voltage and eventually lead to stronger prospective effect than described by our model.

Now we consider threshold adaptation. Similar to before, we start with the output rate, but now defined as a function of the distance of the membrane voltage $v$ to the threshold $\vartheta$, $r = \psi(v - \vartheta) = \psi(v_\vartheta)$, where $v_\vartheta$ describes the voltage relative to the threshold.

The threshold dynamics is such that $\vartheta$ slowly increases as the voltage increases. Specifically, the threshold dynamics is given by
\begin{equation}
    \dot \vartheta = \frac{\vartheta_\infty(v) - \vartheta}{\tau_\vartheta(v)} 
\end{equation}
where the steady state of the threshold $\vartheta_\infty(v)$ is a monotonically increasing function of the voltage (typically a sigmoidal akin to $m_\infty(v)$), and $\tau_\vartheta(v)$ is bell-shaped (in analogy of $\tau_m$ above, but with timescales on the order of $100\,$ms up to $1\,$s).
Analogously as we approximated $h$ above, we now approximate $\vartheta = \vartheta_\infty(v) - \tau_\vartheta(v) \, \dot{\vartheta} \approx \vartheta_\infty(v) - \tau_\vartheta(v) \, \vartheta'_\infty(v) \, \dot v$.
Using this approximation we write the distance to the threshold as

\begin{equation}
    v_\vartheta
    = v - \vartheta
    \approx  v - \vartheta_\infty(v) +  \tau_\vartheta(v) \, \vartheta'_\infty(v) \, \dot v
    = v_\vartheta^\infty(v) + \trthr(v)\,v_\vartheta^{\infty\prime}(v)\,\dot v
    = v_\vartheta^\infty(v) + \trthr(v)\,\dot{v}_\vartheta^\infty(v)
 \label{eq:thr_r}
\end{equation}
with $v_\vartheta^\infty(v)=v-\vartheta_\infty(v)$ the steady state distance between voltage and threshold, and $\trthr = \dfrac{\tau_\vartheta(v)\,\vartheta_\infty'(v)}{v_\vartheta^{\infty\prime}(v)}$ estimating the time constant of the prospective mechanism described by the right-hand side of \cref{eq:thr_r}.

As in the derivation for the prospective firing based on sodium inactivation, we Taylor expand the steady-state function for small variations $\dot v$ as $v_\vartheta^\infty(v + \trthr(v) \, \dot v) \approx v_\vartheta^\infty(v) + \trthr(v) \, v_\vartheta^{\infty\prime}(v) \, \dot v$.
Inserting this approximation into \cref{eq:thr_r}, we rewrite the estimate of the output rate
\begin{equation}
    \rthr
    = \psi(v -\vartheta) 
    \approx \psi( v_\vartheta^\infty(v + \trthr(v)\, \dot v) ) 
    = \psi_\vartheta(v + \trthr(v)\, \dot v) = \rthrpros,
 \label{eq:thr_pros1}
\end{equation}
which implicitly defines the transfer function $\psi_\vartheta$.
As in the Hodgkin-Huxley case, the dynamics of the threshold adaptation is absorbed in the function $\psi_\vartheta$ of the voltage, while the effect of the changing threshold is expressed in the term $\trthr(v)\, \dot v$.
Again, we compare $\rthr$ and $\rthrpros$ to the instantaneous rate $r_\vartheta^\infty(v) = \psi( v_\vartheta^\infty (v) )= \psi_\vartheta^\infty(v)$ as the activation function shown in \cref{fig:5_theory}e leads to skewed rates and makes a comparison to $v$ impractical.

We can also extend our model of prospective coding to active dendrites. Sinusoidal stimulation of apical dendrites {\em in vitro} can lead to an advanced voltage response in the soma of cortical pyramidal neurons \citep{ulrich_dendritic_2002}.
The advanced response is assigned to the dendritic $I_h$ current that slowly counteracts voltage de- and hyper-polarizations in the apical dendrite.
The somatic impact of the dendritic $I_h$ could be described in a 1-compartment model in analogous terms as we described threshold adaptation.
Instead of subtracting a threshold that slowly increases with the voltage, $r= \psi(v - \vartheta)$, one may add an $I_h$ current that slowly decreases with the voltage, $r= \psi(v+I_h)$, leading in a similar way to a temporally advanced response.

\section*{Acknowledgements}
We thank Paul Haider, Kevin Max and Ismael Jaras for enlightening discussions. We gratefully acknowledge funding from the European Union under grant agreement $\# 945539$ (Human Brain Project, SGA3). We would like to express particular gratitude for the ongoing support from the Manfred St\"ark Foundation. Our work has greatly benefited from access to the Fenix Infrastructure resources, which are partially funded from the European Union’s Horizon 2020 research and innovation programme through the ICEI project under the grant agreement No. 800858. This includes access to Piz Daint at the Swiss National Supercomputing Centre, Switzerland.

\section*{Author Contributions}
All authors contributed to the development of the theory and simulations for the spiking neurons, as well as the emergence of prospectivity from negatively coupled adaptation-like variables. SB developed the code and performed the simulations, supervised by KW and FB. SB and FB wrote a first version of the manuscript. All authors contributed to writing the final manuscript. WS developed the theory for the rate-based models and wrote the corresponding sections.

\section*{Competing Interests}
The authors declare there are no competing interests.

\printbibliography{}

\newpage
\appendix
\renewcommand{\thefigure}{SI.\arabic{figure}}
\renewcommand{\thesection}{SI}
\renewcommand{\thesubsection}{SI.\arabic{subsection}}
\section{Supplementary Information}
\setcounter{figure}{0}    

\begin{figure}[H]
    \centering
    \includegraphics[width=1\linewidth]{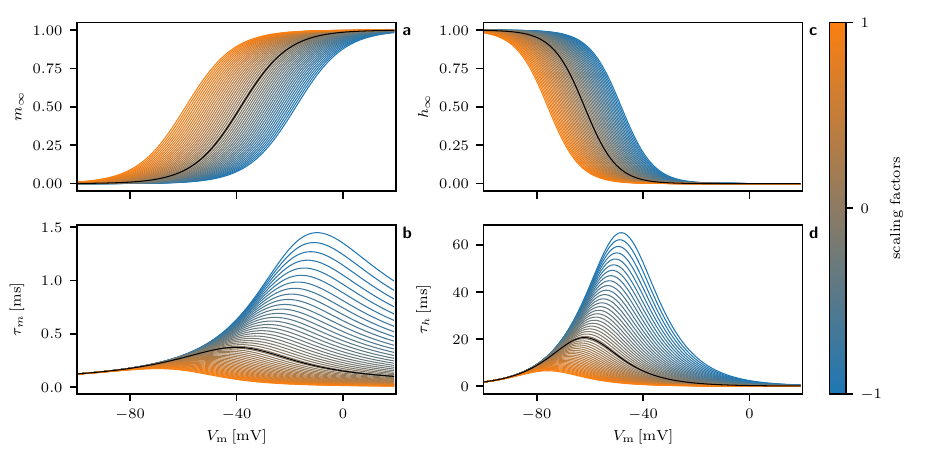}
    \caption{Impact of scaling factors $\lambda$ on the gating variables, where the scaling factors are on a $\log10$ scale resulting in a range from $1/10$ to $10$. {\bf (a, b)} show the same results for the impact of $\bh$ on $\hinf$ and $\tauh$. {\bf (c, d)} show how $\minf$ and $\taum$ are changed when $\am$ is scaled. The black line corresponds to the original variable.}
    \label{fig:A1_scaling_factors}
\end{figure}
\begin{figure}[H]
    \centering
    \includegraphics[width=1\linewidth]{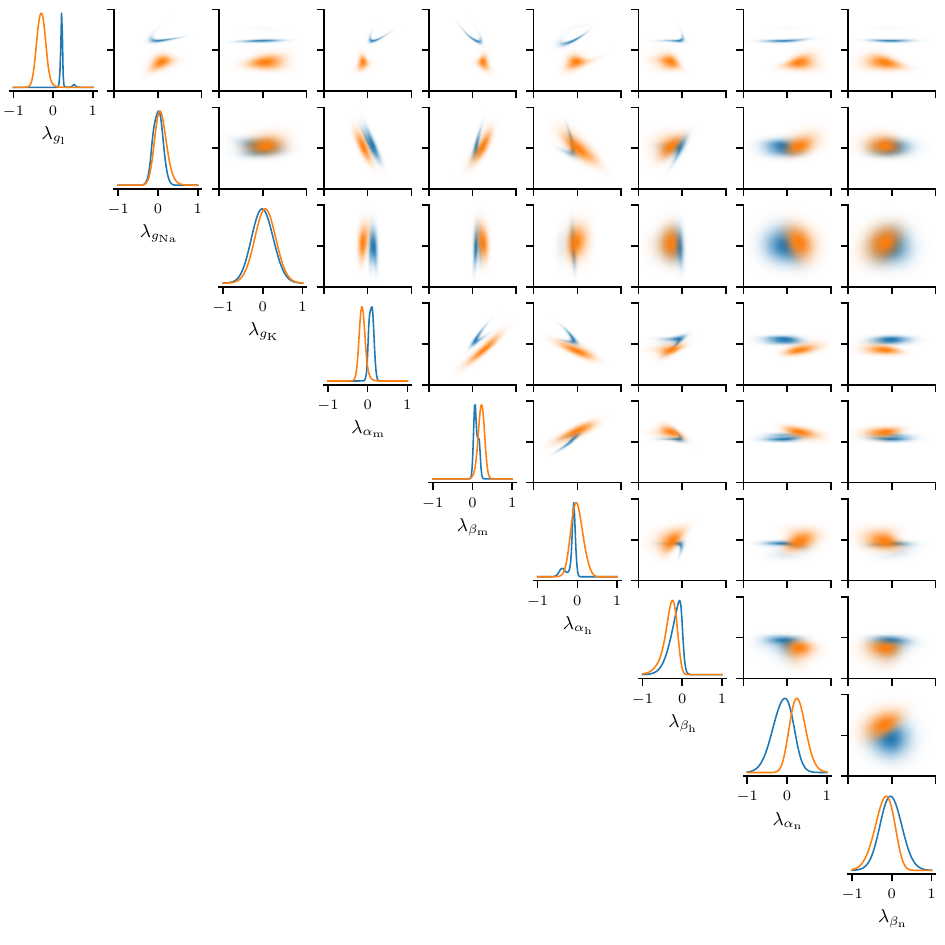}
    \caption{9-dimensional conditional posterior stemming from SBI analysis of the Hodgkin-Huxley model around the cortical parameters. See Methods for a definition of the associated parameters and also \cref{fig:2_posterior} for a detailed description of a similar plot.}
    \label{fig:A2_conditional_posterior_9d}
\end{figure}

% \begin{figure}[H]
%     \centering
%     \includegraphics[width=1\linewidth]{Fig-A3_fitted_gatings.pdf}
%     \caption{With a different notation, the gating variables can be change individually: the steady states can be shifted along the x-axis and the time constants can be scaled along the y-axis. This way, the influence of the steady states and time constants can be analysed independently.
%     {\bf (a, b)} The cortical model of \cref{fig:1_results} as reference.
%     {\bf (c-e)} With a smaller $\taum$ the phase advance gets larger (c), while a larger $\taum$ leads to a smaller advance (e) as compared to the original parameters (a).
%     {\bf (f-h)} The opposite is observed for $\tauh$, with a higher $\tauh$ leading to a larger advance.
%     {\bf (i-k)} The results shifting $\minf$ \SI{2.5}{mV} to the left leading to a smaller phase advance and to the right resulting in a larger phase advance of $\rout$ with respect to $\Iin$ as compared to the original model (a, b).
%     {\bf (l-n)} Shifting $\hinf$ gives the opposite results.}
%     \label{fig:A3_fitted_gatings}
% \end{figure}

\begin{figure}
    \centering
    \includegraphics{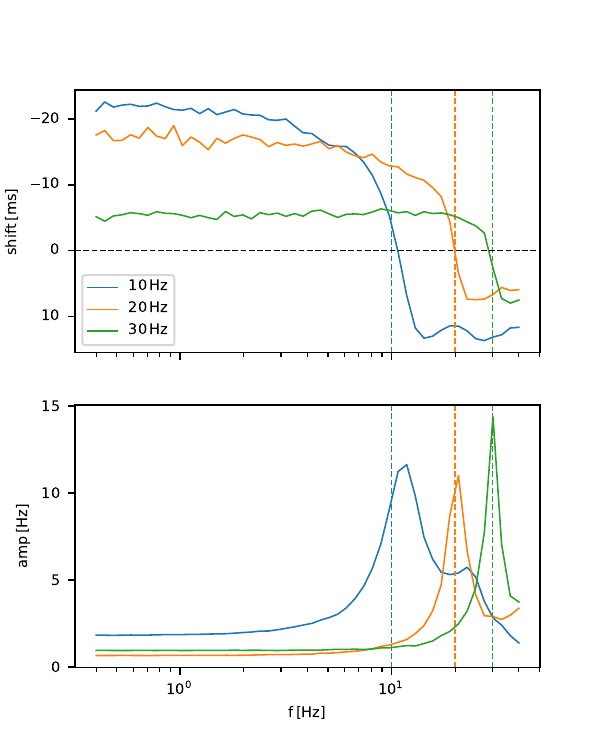}
    \caption{The frequency response of a Hodgkin-Huxley neuron is dependent on the mean of the input current and hence the mean of the output firing rate.
    {\bf (a)} If we increase the mean input current such that the mean output rate of the neuron (dashed vertical line) is larger, the resonance frequency shifts accordingly to a higher frequency, as seen for the amplitude response (solid line).
    We show the frequency response for three output firing rates of \SI{10}{Hz} (blue), as used throughout the manuscript, \SI{20}{Hz} (orange) and \SI{30}{Hz} (green).
    {\bf (b)} The crossing of the phase response with horizontal axis indicating the transition from prospective to retrospective firing shifts together with the resonance to higher frequencies.
    The advance that is observed for lower frequencies gets smaller if we increase the mean input current.
    }
    \label{fig:A4_frequency_response_input_dependency}
\end{figure}

\begin{figure}
    \centering
    \includegraphics{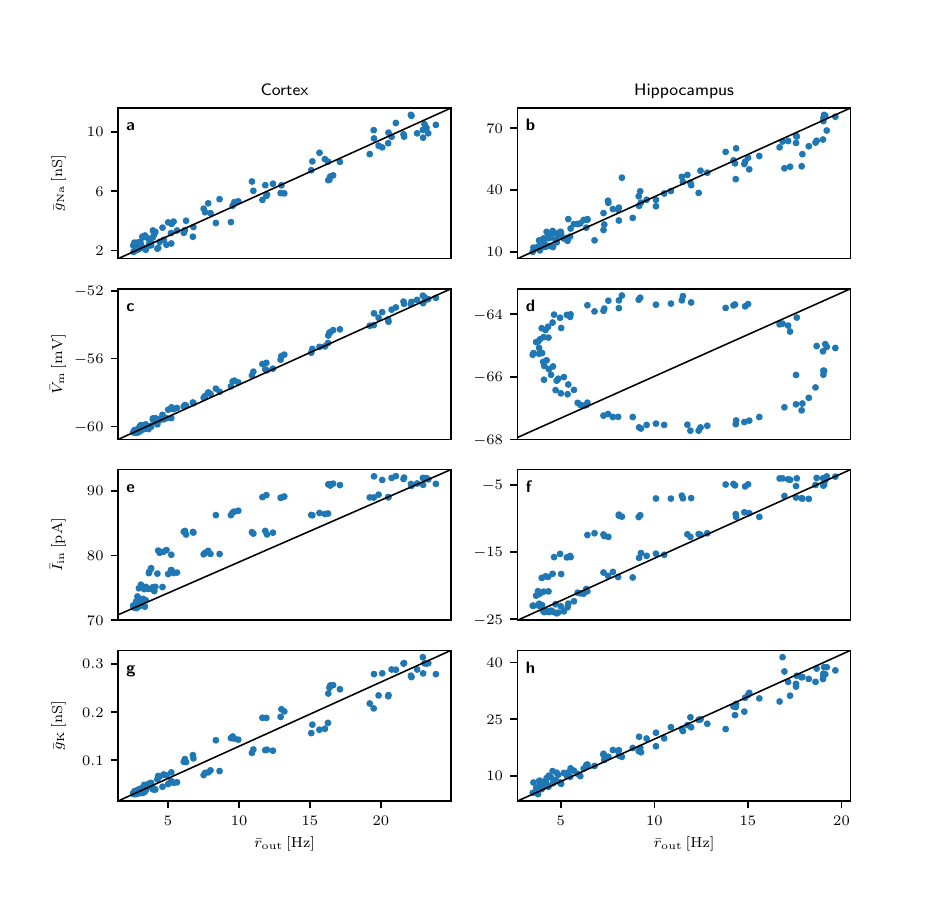}
    \caption{
    {\bf (a, c, e, g)} Hodgkin-Huxley variables plotted versus the output rate for the cortical parameters.
    Points on the dotted line indicate no time shift and proportionality between two variables.
    {\bf (b, d, f, h)} The same plots for the cortical parameters.
    {\bf (a, b)} The sodium conductance is proportional to the output rate for both parameter sets and thus a function of it a good approximation of the output rate of the two neurons.
    {\bf (c, d)} The membrane potential is proportional to the output rate for the cortical parameters (because of the sharp peak of the action potential) but not for the hippocampal parameters (because of strong hyperpolarization).
    {\bf (e, f)} The input current is not proportional for both parameter sets.
    {\bf (g, h)} The potassium conductance is proportional to the output rate of the hippocampal parameters, but not for the cortical parameters.
    }
    \label{fig:A5_r_out_dependency}
\end{figure}

\subsection{Hodgkin-Huxley model}
\begin{align}
   \Cm \dfrac{v(t+\dt) - v(t)}{ \dt} = \gl\,(\El - v) + \gNabar\,m^3\,h\,(E_{\rm Na} - v) + \gKbar\,h^4\,(E_{\rm K} - v) + \Iin\\
    \Iin = I_0 + I_1\,\sin(2\,\pi\,f\,t) + I_{\rm noise}\\
    \dfrac{I_{\rm noise}(t + \dt) - I_{\rm noise}(t)}{\dt} = - \dfrac{1}{\tau_\textrm{noise}}I_{\rm noise}(t) + \sigma\,\sqrt{\dfrac{2}{{\rm d}t\,\tau_{\rm noise}}}\mathcal{N}(0,1)\\
    \dfrac{m(t+\dt) - m(t)}{\dt} = \alpha_m\,(1-m(t)) - \beta_m\,m(t)\\
    \dfrac{h(t+\dt) - h(t)}{\dt} = \alpha_h\,(1-h(t)) - \beta_h\,h(t)\\
    \dfrac{n(t+\dt) - n(t)}{\dt} = \alpha_n\,(1-n(t)) - \beta_n\,n(t)
\end{align}

\subsubsection{Cortical Parameters}
For $v$ in mV
\begin{align}
    \Cm = \SI{1}{\mu F/{cm^2}}\\
    \gl = \SI{3e-4}{S/cm^2}\\
    \El = \SI{-65}{mV}\\
    E_{\rm K} = \SI{-77}{mV}\\
    E_{\rm Na} = \SI{55}{mV}\\
    \gNabar = \SI{40}{mS/cm^2}\\
    \gKbar = \SI{35}{mS/cm^2}\\
    \alpha_n = 0.02\,(v - 25)/(1 - \exp((25 - v)/9))\\
    \beta_n = - 0.002\,(v - 25)/(1 - \exp((v - 25)/9))\\
    \alpha_m = 0.182\,(v + 35)/(1 - \exp(-(35+v)/9))\\
    \beta_m = -0.124\,(v + 35)/(1 + \exp((35+v)/9))\\
    \alpha_h = 0.25\,\exp(-(v+90)/12)\\
    \beta_h = 0.25\,\exp((v+62)/6)/\exp((v+90)/12)
\end{align}

\subsubsection{Hippocampal parameters}
Again, $v$ in mV
\begin{align}
    \Cm = \SI{1}{\mu F/{cm^2}}\\
    \gl = \SI{5e-5}{S/cm^2}\\
    \El = \SI{-60}{mV}\\
    E_{\rm K} = \SI{-90}{mV}\\
    E_{\rm Na} = \SI{50}{mV}\\
    \gNabar = \SI{100}{mS/cm^2}\\
    \gKbar = \SI{30}{mS/cm^2}\\
    v_T = \SI{-63}{mV}\\
    \alpha_m = 0.32\,(13-v+v_{\rm T})/(\exp((13-v+v_{\rm T})/4)-1)\\
    \beta_m = 0.28\,(v-v_{\rm T}-40)/(\exp((v-v_{\rm T}-40)/5) - 1)\\
    \alpha_h = 0.128\,\exp((17-v+v_{\rm T})/18)\\
    \beta_h = 4/(1+\exp((40-v+v_{\rm T})/5))\\
    \alpha_n = 0.032\,(15-v+v_{\rm T})/(\exp((15-v+v_{\rm T})/5) - 1)\\
    \beta_n = 0.5\,\exp((10-v+v_{\rm T})/40)
\end{align}

\end{document}